# Factors Influencing Farmers' Motivation to Adopt Smart Farm Technology in South Korea

Jihyuk Bang, Ji Woo Han

**[Abstract]**

      Smart farming technologies have recently become a major focus because they promise improved agricultural productivity together with sustainability benefits. The rate at which farmers adopt new technologies differs because of various socio-economic and technological factors. This research investigates the main factors affecting smart farm adoption through an evaluation of farmer age, education attainment, land size, government banking, technological hurdles and financial limitations. Statistical analysis of survey responses reveals that farmers under thirty years old and possessing advanced education along with operating expansive farms demonstrate greater willingness to adopt smart farming solutions. Government policies that offer financial assistance and training programs drive farmers to adopt new systems yet technical and financial difficulties prevent broad adoption. The research results demonstrate the requirement for targeted policies that should help farmers by providing specific financial help and digital education programs. This study delivers important knowledge which helps government officials together with agricultural leaders to accelerate the adaptation speed of smart farming technology.

# 1. [Introduction]

## 1.1 Background

The adoption of smart farm technology represents a worldwide solution which addresses agricultural challenges between environmental impact and productivity and efficiency. Multiple types of technologies form the basis of this system including automated agricultural management systems, agricultural drones, automated unmanned tractors using AI, remote management sensors and data analysis. Agriculture 4.0, which includes smart farming technologies, is considered to have tremendous potential to be identified and utilized due to its ease of operation and cost reduction (Javaid et al., 2022). The widespread adoption and application of smart agriculture on farms has beneficial effects on agricultural management and economics.

South Korea has witnessed significant advancements in smart farm technologies. The country's government and agricultural institutions have actively promoted the adoption of smart farm technologies, recognizing their potential to enhance agricultural sustainability and competitiveness. Smart Farm Korea, a smart farm manpower training center hosted by a national institution, is making efforts to educate and support people on smart farm technology on a regular basis (Smart Farm Korea, 2023). However, despite the availability of these technologies, data suggest that the rate of adoption and integration of these technologies by farmers in Korea is 2.8% lower than the rate of smart farm technology adoption by farmers in all countries around the world (Byun, 2022).

## 1.2 Statement of the Problem

Farmer acceptance of smart farm technologies depends on various elements that encompass both motivational aspects and technological understanding as well as demographic features (Annosi et al., 2019). The knowledge of farmers' motivational factors regarding technology adoption proves essential because it reveals adoption obstacles and increases economic performance and environmental sustainability while promoting educational sharing and guiding policy improvements. New discoveries about these agricultural sector obstacles will help decrease slow implementation rates and economic waste along with environmental damage and information gaps and insufficient policy backing.

The growing significance of smart farming receives limited academic attention regarding farmers' motivation levels and their adoption behavior of smart farm technologies specifically within South Korea. This study lacks sufficient investigation into how social factors influence the relationship between motivation and smart farm technology adoption. The analysis of motivation factors for smart farm technology adoption requires deeper investigation because South Korea faces specific agricultural challenges of scarce arable land alongside an aging farm workforce and government-led technological programs.

This research tries to connect this knowledge gap by studying farmers' motivation toward smart farm technology adoption and their demographic characteristics. This study intends to deliver essential findings which will help policymakers along with researchers and agricultural practitioners to facilitate successful smart farming technology adoption and integration in South Korea.

**1.3 Purpose Statement**

This research investigates how Rogers' (2003) Innovation Diffusion Theory explains the connection between South Korean farmers' smart farming technology adoption motivation and actual technology adoption rates. The research examines demographic elements which function as potential variables that affect the process.

The independent variable expresses farmers' interest and proactive adoption behavior toward integrating advanced farm technologies in their agricultural operations. The dependent variable measures how many farmers successfully execute smart farming systems through implementation.

The analysis evaluates how demographic traits such as age, gender, income levels and educational attainment affect both why farmers adopt smart farm technologies and their actual adoption rates. Such factors help researchers understand the ways different types of farmers affect their adoption behavior.

This research investigates the main motivational factors which drive farmers to use smart farming technologies through an analysis of technological preparedness along with economic factors and personal qualities. This research details the essential factors to provide knowledge that helps policy makers create better strategies for technology adoption and sustainable agricultural growth in South Korea..

2. [Literature Review]

**2.1 Farmers' Motivation and Innovation Diffusion in Smart Farm Technology Adoption**

This literature review covers the complex process of adopting smart farm technologies by Korean farmers, which is influenced by various variables, including farmers' motivation. However, the relationship between farmers' motivation and the adoption rate of smart agricultural technology in Korea has received minimal attention in previous studies. In the case of South Korea, the motivations for farmers to adopt smart farming technologies are influenced by several unique characteristics and the agricultural landscape of the country. These include limited land availability, labor shortage, an aging population, precision agriculture for resource optimization, climate and weather challenges, government support and initiatives, and market competitiveness and consumer demand. These motivations reflect the specific agricultural context and priorities of South Korea. While the general benefits of smart farming technologies apply, the unique challenges and opportunities in the country shape the motivations for farmers to adopt these technologies. Through our literature review, we have discovered that the adoption of smart farm technologies by Korean farmers is a complex process influenced by various factors. The motivations discussed above play an important role in determining the adoption rate of these technologies. However, we believe that further research is needed to further understand the relationship between farmers' motivation and the adoption of smart farming technologies in Korea.

Innovation Diffusion Theory by Rogers (2003) has been applied to a variety of fields. This theory shows how innovations are adopted and diffused within a social system. The five factors that influence an individual's acceptance of innovation that Rogers identified are relative advantage, compatibility, complexity, trialability, and observability. In the context of adopting smart farm technologies, Yoon et al. (2020) examined three factors from Roger's theory, relative advantage, complexity, and compatibility. They applied financial costs, the transition to the digital environment, and technological compatibility for the variables to address these three diffusion factors. Trialability and observability were excluded from their research, following the IT adoption model (Tornatzky & Klein, 1982).

**2.2 Compatibility: Influence of Characteristics to Adopting Technology**

From the study by Lowenberg-DeBoer & Erickson (2019), the reason why the adoption of precision agriculture (PA) has been slow is looked into. A meta analysis was used to collect information on adoption rate of PA and the farmers' interest in these technologies. They found that in North America, farm size, types of crop being cultivated, farmers working in PA workshops, prior interest into technologies, and younger and high revenues were related to the high interest and adoption rate of PA. This shows that farmers with those characteristics are more likely to be motivated to adopt smart farm technologies. This research found that the level of mechanical systems is positively correlated with the adoption

(willingness to adopt) of PA. From Roger's innovation theory, complexity can be applied to these variables.

From this article (Lowenberg-DeBoer & Erickson, 2019), the variable called the level of mechanical systems is reflected in the research as a necessary factor in adopting smart farm technology to Korean farmers. In addition, Roger's complexity factor and this variable are related to the system analysis of smart farm technology adoption. In the context of smart farm technology, complexity can be linked to the technological readiness of farmers. Farmers' level of technological readiness, including familiarity with and ability to use alcohol, influence perceptions of complexity. The perceived complexity of smart farm technologies can impact farmers' motivation to adopt them. Based on these characteristics, additional research is needed to apply to Korean farmers' motivations for adopting smart farm technology.

**2.3 Relative Advantage: Economic Factors**

The Theory of Economic Viability represents a widely acknowledged concept within agricultural economics research about technology adoption. No single originator or source has established this concept but it describes how smart farming technology adoption leads to economic benefits for farmers. The theory shows that advanced farming techniques together with precision technology enable farmers to improve their productivity while decreasing costs and unlocking better market access to bring profitable sustainability to their agricultural business. The theory supports agricultural economic efficiency standards which promote optimization throughout the agricultural sector. Kutter et al. (2011) analyzed precision farming adoption across Germany and the Czech Republic and Denmark and Greece and discovered economic factors are the main drivers behind precision farming practice adoption. Large farms use precision farming methods to boost their productivity and minimize costs and improve their market connections. The research shows that joint ventures and cooperation in precision farming machinery aligns with the Theory of Economic Viability since collaborative efforts enable shared costs and increased resource access and enhanced efficiency. Farmers need to work together through cooperative initiatives to access the benefits of smart farming technologies and make their agricultural operations economically sustainable. The study by Kutter et al. demonstrates how smart farming technologies create economic sustainability improvements throughout the agricultural industry.

Economic analyses have demonstrated the significant production-inducing effects and value-added effects of smart farm technologies on the agricultural market and various industries (Ramil, Daely, Kim, & Lee, 2020). The study from O'Shaughnessy, Kim, Lee, Kim, Kim, and Shekailo conducted an economic feasibility analysis, showing the viability of smart farm systems in tomato and strawberry farming and providing insights into adoption factors and transition rates (O'Shaughnessy et al., 2021). From Roger's innovation diffusion theory, relative advantage, compatibility, and complexity can be applied.

**2.4 Definition of Terms**

Smart Farm: Smart farms are farms that use advanced and new technologies such as big data, internet of things (IoT), precision agriculture, and more that would benefit the farming practices in the agricultural sector.

- Smart Farmers: Refers to farmers using smart farming technologies.
- Traditional Farm: Farms that do not use smart farm technologies, but use traditional farming methods.
- Traditional Farmers: Refers to farmers using traditional farming methods.
- Motivation: Motivation is the willingness of an individual to do something. In this study, motivation refers to the farmers' willingness to implement smart technologies.

- Characteristics: Characteristics refers to a feature or a quality that distinguishes an individual, such as demographic factors like age, gender, occupation, to general characteristics such as farm size, number of employees, and farming experiences.
- Economic Factors: Economic factors refer to the various conditions and forces that influence the production, distribution, and consumption of goods and services within an economy.

This literature review explores the adoption of smart farm technologies by Korean farmers, focusing on the influence of factors such as farmers' motivations. The motivations in South Korea are shaped by characteristics such as limited land availability, labor shortage, an aging population, precision agriculture for resource optimization, climate and weather challenges, government support and initiatives, and market competitiveness and consumer demand. The literature review highlights the complex nature of the adoption process and the importance of understanding farmers' motivations. Moreover, it discusses the application of Rogers; Innovation Diffusion Theory and the factors, compatibility, complexity, and relative advantage.

'Table 1 Summary of findings' is composed of variables from research articles constituting the literature review, and evaluates influences on the adoption of smart farm technology by farmers. It is the result of a process for listing effects positive or negative.

### 2.4.1 Table 1. Summary of Findings

| I.V. | Reference | Effects | D.V. | Comments |
|---|---|---|---|---|
| Age | (Annosi et al., 2019) | (-) | Willingness of adopting 4.0 tech | Take Compatibility |
| Farm Size | (Annosi et al., 2019) | (+) | Willingness of adopting 4.0 tech | Take Compatibility |
| Education Level | (Annosi et al., 2019) | (+) | Willingness of adopting 4.0 tech | Take Compatibility |
| Farming Experience | (Annosi et al., 2019) | (+) | Willingness of adopting 4.0 tech | Take Compatibility |

| Number of Employees | (Annosi et al., 2019) | (+) | Willingness of adopting 4.0 tech | Take Compatibility |
|---|---|---|---|---|
| Gender | (Abdollahzadeh et al., 2016) | (+): Males are more likely to adopt 4.0 tech | Willingness of adopting 4.0 tech | Take Compatibility |
| Possibility of Reducing Costs (Cost Effect) | (Kutter et al., 2011) | (+): While costs of fertilizer and pesticides (cost of production) will reduce (-): the cost of equipment, maintenance and learning curve increases | Willingness to adopt PF tech | Take, can be used for economic factors |
| Joint Investment | (Kutter et al., 2011) | (+): Can teach each other to reduce learning costs and have compatibility: Can teach each other to reduce learning costs and have compatibility within the same tech system | Willingness to adopt PF tech | Take, can be used for economic factors |
| Interest in Technology | (Lowenberg-DeBoer & Erickson, 2019) | (+) | Willingness to adopt PA | Take, Compatibility |
| Market Access | (Kutter et al., 2011) | (+) | Willingness to adopt PF tech | Take, Relative Advantage |
| Productivity | (Kutter et al., 2011) | (+) | Willingness to adopt PF tech | Take, Relative Advantage |
| Production-inducing Effects | (Ramil, Daely, Kim, & Lee, 2020) | (+) | Willingness to adopt SF | Take, Relative Advantage |
| Value-added Effects | (Ramil, Daely, Kim, & Lee, 2020) | (+) | Willingness to adopt SF | Take, Relative Advantage |

## 2.4.2 Figure 1. Path Diagram for the Research Model

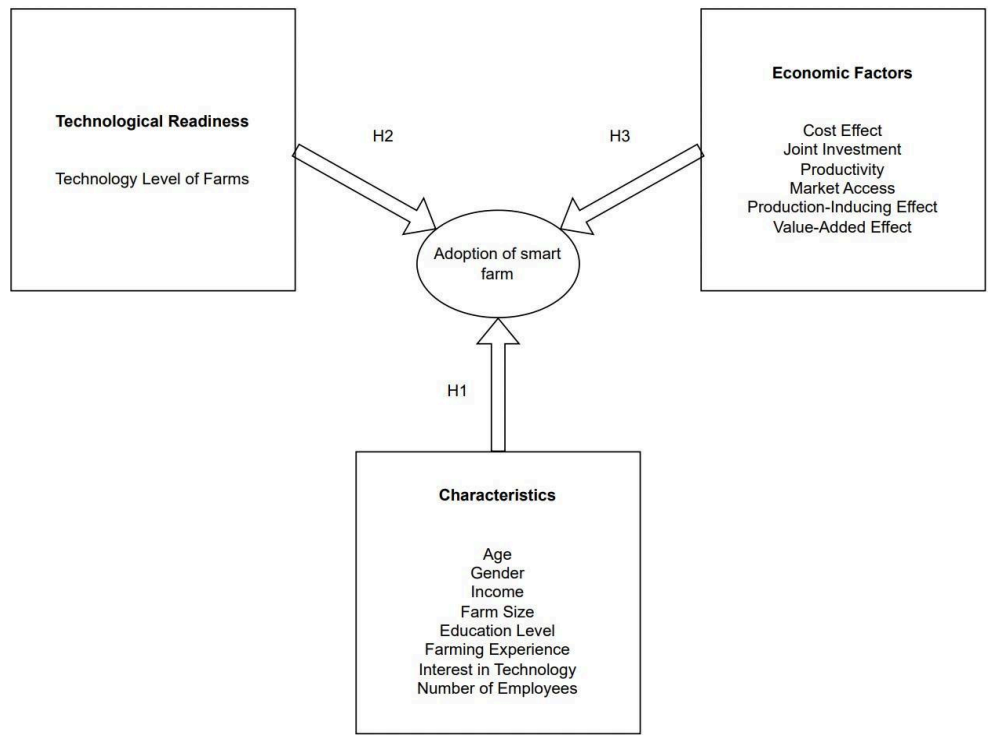

## 3. [Research Framework]

### 3.1 Research Question

Although the adoption of smart farm technologies has much potential in the agricultural sector, there is still a need to understand what drives the farmers to be motivated to adopt them, particularly in Korea. These research questions and hypotheses aim to examine how the motivations to adopt smart farm technologies differ based on the characteristics of farmers and the economic and technological factors that influence their decision-making.

1. How do these motivations differ based on the characteristics of farmers such as age, gender, education, income, and farm size?
2. What are the economic and technological motivations for individual farmers to adopt smart farm technology in Korea?

**3.2 Research Model**

As illustrated from Figure 1, our research model shows the relationship between the various variables, including the characteristics of farmers, technological readiness, economic factors, and the adoption of smart farm technologies. This model looks at the influence of the different characteristics, technological readiness, and economic considerations on farmers' motivation to adopt smart farm technologies. By examining these factors, we could understand what drives the adoption of smart farming technologies in Korea.

**3.3 Hypothesis**

Hypothesis 1: According to Annosi et al. (2019), there is a positive relationship between the characteristics (age, gender, income, education, farm size, farming experience, interest in technology, and number of employees) and farmers' motivation to adopt smart farm technology. So in our research, we are going to examine the relation as this:

- Alternative hypothesis (H1): Demographic factors have a significant effect on farmers' motivation to adopt smart farm technologies.
    - Demographic factors will be measured with age, gender, income, education, farm size, farming experience, interest in technology, and number of employees.

Hypothesis 2: According to Lowenberg-DeBoer & Erickson (2019), there is a positive relation between the technological readiness of farms and farmers' motivation to adopt smart farm technology. (Korean case?, green house) So in our research, we are going to examine the relation as this:

- Alternative hypothesis (H2): Farmers with higher technological readiness exhibit higher motivation to adopt smart farm technologies.

Hypothesis 3: According to Kutter et al. (2009), there is a positive relation between the economic factors and farmers' motivation to adopt smart farm technology. So in our research,

we are going to examine the relation as this:

- Alternative hypothesis (H3): Farmers are more motivated to invest in smart farm technologies due to the potential economic benefits and efficiency improvement.
    - Economic benefits will be measured with this cost effect, value-added effects, market access, and joint investment.
    - Efficiency improvement will be measured with productivity and production-inducing effects.

4. **[Methods and Design]**

**4.1 Methodology**

This quantitative study will examine the farmers' motivations for smart farm technology adoption in Korea. We chose to conduct a quantitative research due to the lack of quantitative studies on the motivations of adopting smart farming technologies in Korea. This method is appropriate for this research because it allows for the measurement and analysis of farmers' motivations and demographic characteristics in a structured manner. The target population of this research will be farmers in both smart farms and traditional farms in Korea.

**4.2 Sampling Strategy**

Out of 1,022,797 farming households, about 11% use smart farming technologies (Agricultural, Forestry, and Fisheries Survey, 2023; Byun, 2022). Random sampling will be used to select participants for traditional farmers, and purposive sampling techniques will be used to select participants for smart farmers as the proportion of smart farmers are relatively small (11%) (Byun, 2022). A stratified sampling method will be used by conducting information from 100 participants, where 89 will be traditional farmers and 11 will be smart farmers for a proportional number of participants from each group based on the actual distribution of smart farmers.

Collaborating with the local agricultural offices that have direct access to farmers in Korea is a method to gain the contact information of farmers to distribute the survey. We will also utilize Nonghyup Online Market, which is an online market platform designed for farmers to connect directly with their consumers. After seeking permission from the platform administrations, we will distribute a brief description of the study and a link to the online survey, where interested farmers can voluntarily provide with their contact details. Sample questionnaires are provided in Appendix.

### 4.3 Analysis Strategy

Our research data will be gathered through questionnaires which survey the selected participants. Our survey questionnaire sample will be built and collected through SurveyMonkey. SPSS (Statistical Package for the Social Sciences) will serve as the instrument for analyzing the gathered data. The research will establish a significance level of 0.05 (5%) while setting a desired power level at 0.80 (80%) with an estimated correlation of 0.5 for moderate effect since no previous literature exists to estimate. The power analysis shows that at least 30 participants are needed for achieving the target power level.

### 4.4 Ethical Considerations

Research integrity along with participant privacy received protection as all respondents signed consent forms before joining. The study maintains full anonymity while following ethical procedures that restrict data use to academic research only.

### 4.5 Limitations

The main drawback of this research depends on outdated data usage. The extended research period creates possible inconsistencies because technological progress and agricultural sector transformations might have taken place during this period. The research results and conclusions from this study may not accurately represent the current conditions regarding farmers' motivation and their use of smart farm technologies.

The research faces an obstacle because of possible language difficulties. The availability of data in Korean language alone creates a potential risk to omit important studies or sources which were conducted in different languages. The study's data selection and synthesis process might develop a potential bias because of language barriers which could reduce the study's comprehensive and inclusive findings.

5. [Conclusion]

The proposed research aims to investigate the motivations for smart farm technology adoption among farmers in Korea. This study aims to fill the research gap, where motivations were studied on a macro level and considering the farmers' personal motivation may shed a different light, through investigating the relationship between farmers' motivation to adopt smart farm technologies in Korea based on the influence of various factors. By applying the three factors from the Innovation Diffusion Theory by Rogers, which are relative advantage, complexity, and compatibility, we will explore the connection between farmers' motivation to adopt smart farm technologies. The findings will have implications for policymakers, agricultural practitioners, and researchers in promoting the widespread implementation of smart farm technologies in Korea. The research methodology involves quantitative analysis using questionnaires to gather data on farmers' motivations based on characteristics, technological readiness, and economic motivations.

This study will contribute to the field by providing valuable insights to the related field or audiences. Policymakers or government agencies in Korea can use this study to create programs that could promote the adoption of smart farming technologies. Farmers could also use this study to gain practical insights in their agricultural practices when implementing smart farming.

Moreover, this study will enrich the existing literature on the factors influencing the adoption rate of smart farming, especially contributing to the Korean agricultural context.

## 6. [Appendix]

Survey Sample

### Q1. Demographic/Characteristic Factors

**What is your age?**

- Under 30
- 30-40
- 41-50
- 51 and above

**What is your gender?**

- Male
- Female
- Prefer not to say

**What is your level of education?**

- High school or below
- Associate's degree
- Bachelor's degree
- Master's degree or higher

**What is your annual income?**

- Less than $20,000
- $20,000-$40,000
- $40,000-$60,000
- More than $60,000

**What is the size of your farm?**

- Less than 1 hectare/acre
- 1-5 hectares/acres
- 6-10 hectares/acres
- More than 10 hectares/acres

**How many years of farming experience do you have?**

- Less than 5 years
- 5-10 years
- 11-20 years
- More than 20 years

**How would you rate your interest in technology?**

- (1 = No Interest, 10 = Very Interested)

---

**Q2. Technological Readiness Motivation**

**Please indicate the technological readiness of your farm in terms of adoption of modern agricultural technologies:**

- **Low**: Limited or no adoption of modern agricultural technologies
- **Medium**: Partial adoption of modern agricultural technologies
- **High**: Extensive adoption of modern agricultural technologies

**On a scale of 1 to 10, how satisfied are you with the stability of the network connection for smart farm technologies?**

- (1 = Not important at all, 10 = Extremely important)

**Have you experienced any network errors or issues while using smart farm technologies?**

- Yes
- No

**On a scale of 1 to 10, how important is network stability for the successful implementation of smart farm technologies in your opinion?**

- (1 = Not important at all, 10 = Extremely important)

**Which of the following factors influence network stability in your farm? (Select all that apply)**

- Weak signal strength
- Interference from other devices
- Distance from network access points
- Technical issues with smart farm devices
- Other (please specify)

**How would you rate the quality and accessibility of the database used for smart farm technologies?**

- Poor
- Fair
- Good
- Excellent

**Have you encountered any challenges or limitations in accessing relevant data for smart farming?**

- Yes
- No

**Do you believe that farm size influences the adoption rate of smart farm technologies?**

- Yes
- No

---

## Q3. Economic Motivation

**In your opinion, how much do smart farm technologies contribute to increasing profits and economic efficiency in farming?**

- Not at all
- Slightly
- Moderately
- Significantly

**Have you observed any noticeable improvements in your financial performance since implementing smart farm technologies?**

- Yes
- No

**How would you rate the cost-effectiveness of adopting smart farming technologies compared to traditional farming methods?**

- (1 = Not important at all, 10 = Extremely important)

**Have you collaborated with other farmers or agricultural organizations to jointly invest in smart farm technologies?**

- Yes
- No

**If you answered "Yes" for the previous question, how would you rate the overall benefits from the joint investment?**

- (1 = No Benefits, 10 = Extreme Benefits)

**Do you believe that smart farm technologies have improved the overall efficiency of your farming experience?**

- Yes
- No

---

These survey questions can help gather information about farmers' motivations, demographic factors, technological readiness, and economic motivations.

# References

Abdollahzadeh, G., Sharifzadeh, M. S., & Damalas, C. A. (2016). Motivations for adopting biological control among Iranian rice farmers. *Crop Protection, 80*, 42–50. https://doi.org/10.1016/j.cropro.2015.10.021

Annosi, M. C., Brunetta, F., Monti, A., & Nati, F. (2019). Is the trend your friend? An analysis of Technology 4.0 investment decisions in agricultural SMEs. *Computers in Industry, 109*, 59–71. https://doi.org/10.1016/j.compind.2019.04.003

Byun, J. (2022). *Analysis on the status and future development of smart farming projects* (In Korean). National Assembly Budget Office. https://www.nabo.go.kr/common/download.jsp?fCode=33317036&fName=%EC%8A%A4%EB%A7%88%ED%8A%B8%EB%86%8D%EC%97%85+%EC%9C%A1%EC%84%B1%EC%82%AC%EC%97%85+%EC%B6%94%EC%A7%84%ED%98%84%ED%99%A9%EA%B3%BC+%EA%B0%9C%EC%84%A0%EA%B3%BC%EC%A0%9C.pdf

Choi, H. (2020). A study on the change of farms using artificial intelligence: Focused on smart farms in Korea. *Journal of Physics: Conference Series, 1642*, 012025. https://doi.org/10.1088/1742-6596/1642/1/012025

Javaid, M., Haleem, A., Singh, R. P., & Suman, R. (2022). Enhancing smart farming through the applications of Agriculture 4.0 technologies. *International Journal of Intelligent Networks, 3*, 39–57. https://doi.org/10.1016/j.ijin.2022.10.004

Kim, D., & Jin, S. (2022). Innovation capabilities and business performance in the smart farm sector of South Korea. *Journal of Open Innovation: Technology, Market, and Complexity, 8*(4), 204. https://doi.org/10.3390/joitmc8040204

Kim, S., Lee, M., & Shin, C. (2018). IoT-based strawberry disease prediction system for smart farming. *Sensors, 18*(11), 4051. https://doi.org/10.3390/s18114051

Korea Statistical Information Service (KOSIS). (2023). *Agricultural, forestry, and fisheries survey* (In Korean). Statistics Korea. Retrieved July 13, 2023, from https://kosis.kr/statHtml/statHtml.do?orgId=101&tblId=DT_1EA1011

Kutter, T., Tiemann, S., Siebert, R., & Fountas, S. (2011). The role of communication and cooperation in the adoption of precision farming. *Precision Agriculture, 12*(1), 2–17. https://doi.org/10.1007/s11119-009-9150-0

Lee, W., & Kim, H. (2020). Assessing the adoption potential of a smart greenhouse farming system for tomatoes and strawberries using the TOA-MD model. *Korean Journal of Agricultural Science, 47*(3), 351–362. https://doi.org/10.7744/kjoas.20200047

Li, J., Liu, G., Chen, Y., & Li, R. (2023). Study on the influence mechanism of adoption of smart agriculture technology behavior. *Scientific Reports, 13*, 10245. https://doi.org/10.1038/s41598-023-35091-x

Lowenberg-DeBoer, J., & Erickson, B. (2019). Setting the record straight on precision agriculture adoption. *Agronomy Journal, 111*(4), 1552–1569. https://doi.org/10.2134/agronj2018.12.0779

O'Shaughnessy, S. A., Kim, M., Lee, S., Kim, Y., Kim, H., & Shekailo, J. (2021). Towards smart farming solutions in the U.S. and South Korea: A comparison of the current status. *Geography and Sustainability, 2*(4), 312–327. https://doi.org/10.1016/j.geosus.2021.12.002

Ramli, M. M., Daely, P. T., Kim, D., & Lee, J. S. (2020). IoT-based adaptive network mechanism for reliable smart farm systems. *Computers and Electronics in Agriculture, 170*, 105287. https://doi.org/10.1016/j.compag.2020.105287

Smart Farm Korea. (2023). *Smart Farm Korea website*. Retrieved from https://www.sfarm.or.kr

Tornatzky, L. G., & Klein, K. J. (1982). Innovation characteristics and innovation adoption-implementation: A meta-analysis of findings. *IEEE Transactions on Engineering Management, 29*(1), 28–45. https://doi.org/10.1109/TEM.1982.6447463

Yoon, C., Lim, D., & Park, C. (2020). Factors affecting adoption of smart farms: The case of Korea. *Computers in Human Behavior, 108*, 106309. https://doi.org/10.1016/j.chb.2020.106309